\documentclass[preprint]{emulateapj}

\slugcomment{Accepted for Publication in the Astrophysical Journal}
\shortauthors{Zaritsky, Zabludoff, \& Gonzalez}
\shorttitle{Globular Clusters and the FM}

\begin{document}
\title{Star Clusters, Galaxies,  and the Fundamental Manifold}
  
\author{Dennis Zaritsky\altaffilmark{1}, Ann I. Zabludoff\altaffilmark{1}, \& Anthony H. Gonzalez\altaffilmark{2}}
\altaffiltext{1}{Steward Observatory, University of Arizona, 933 North Cherry Avenue, Tucson, AZ 85721}
\altaffiltext{2}{Department of Astronomy, University of Florida, Gainesville, FL 32611}

\email{dzaritsky@as.arizona.edu, azabludoff@as.arizona.edu, anthony@astro.ufl.edu}

\begin{abstract}    
We explore whether global observed properties, specifically half-light radii, mean surface brightness, and integrated stellar kinematics,  suffice to unambiguously differentiate galaxies from 
star clusters, which  presumably formed differently and lack dark matter halos.
We find that star clusters 
lie on the galaxy scaling relationship referred to as the Fundamental Manifold (FM), on the extension of a sequence of compact galaxies, and
so conclude that there is no simple way to differentiate 
star clusters from ultra-compact
galaxies. 
By extending the validity of the FM over a larger range of parameter
space and a wider set of objects, we demonstrate that the physics that 
constrains the resulting baryon and dark matter distributions in stellar systems is more
general than previously appreciated.
The generality of the FM
implies 1) that the stellar spatial distribution and kinematics of one type of stellar system do not
arise solely from a process particular to that
set of systems, such as violent relaxation for elliptical galaxies, but are instead
the result of an interplay of all processes responsible for 
the generic settling of baryons in gravitational potential wells, 2) that the physics of how
baryons settle is independent of whether the
system is embedded within a dark matter halo, and 3) that peculiar initial conditions at formation or
stochastic events during evolution do not ultimately disturb the 
overall regularity of baryonic settling.
We also utilize the relatively simple nature of star clusters to relate deviations from the FM to the age of the stellar population and find that stellar population models systematically and significantly over predict the mass-to-light ratios of old, metal-rich clusters. 
We present an empirical calibration of stellar population mass-to-light ratios with age and color.   Finally, we use the FM to estimate velocity dispersions for the  low-surface brightness, outer halo clusters that lack such measurements. 

\end{abstract}


\section{Introduction}
\label{sec:intro}

Stellar systems are conventionally divided into two broad categories, stellar clusters and
galaxies, depending on whether they lie within a dark matter halo. Yet this scheme faces 
conceptual and practical difficulties.  For example, the dynamical masses of certain
``galaxies" \citep[ultra-compact dwarf galaxies or UCDs;][]{mieske} suggest little or no dark matter, as does the modeling
of tidal dwarfs \citep{barnes}. Furthermore, even for indisputable galaxies like normal 
ellipticals, it can be difficult to infer dark matter from kinematic measurements within the 
luminous radius \citep{cappellari}.
It is the spatial concentration of baryons relative to dark matter within the luminous region that distinguishes UCDs
from other galaxies and that makes dark matter easier to measure in spirals than ellipticals.  Rather than differentiating stellar systems with a binary classification --- dark matter halo vs. no halo --- that is difficult to measure,
we consider baryon vs. dark matter concentration, which is easier to observe and allows for a continuum of properties.

Our previous work on scaling relations \citep{zgz, zgzb, zzg} shows that galaxies lie on a progression of star formation efficiencies and baryon concentrations within dark matter halos.  
The mass-to-light ratio $\Upsilon_e$ within the half-light radius $r_{e}$ is one gauge of the relative
concentration of baryons to dark matter.  The large $\Upsilon_e$'s of 
dwarf spheroidals (dSphs) reflect the importance of dark matter within their luminous region, while the lower $\Upsilon_e$'s of ellipticals arise from the dominance of the baryons there.
One could hypothesize a more extreme case than that for ellipticals, where the baryons are packed so tightly that the dark matter is dynamically negligible within $r_{e}$.  It would be difficult to differentiate such a low $\Upsilon_e$ galaxy, if it were low mass, from a star cluster, which is devoid of dark matter.   If they have any dark matter, UCDs \citep{mieske} may be examples of galaxies
whose spatial and kinematic structure thus resemble those of star clusters.  
Are there other observables that reveal their presumably distinct formation and evolutionary paths?
Remarkably, comparably low mass systems can have low $\Upsilon_e$ (stellar clusters, UCDs) or high $\Upsilon_e$ (dSphs). What variations in dark halo properties
of baryonic processes explain this wide range of baryon to dark
matter concentrations?

Such questions are fundamental to understanding 
the nature of low mass stellar systems,
the basic characteristics of galaxies,
and the ways in which baryons can settle into a dark matter halo.  
Here we use a new scaling relation (the Fundamental Manifold of \citet[FM;][]{zgz,zzg})
to explore the first question above, whether
star clusters can be distinguished from low mass galaxies with similar $\Upsilon_e$'s and the implications for their evolution.

In certain projections of parameter space, globular clusters are distinct
from galaxies \citep{kormendy, burstein}.  (Then again, in certain projections, dwarf galaxies
are distinct from giant galaxies.)  Projections of distributions in 
multi-dimensional parameter space can be misleading. It is critical that the defined 
parameter space 1) has  
physically meaningful axes, so that one can interpret the results, 2) produces a galaxy distribution as thin as that allowed by observational errors (i.e., that has $N-1$ dimensionality), thus ensuring that the full dimensionality of the distribution is captured, and 3) exploits the full range of galaxy properties to maximize the discriminatory power of the observational space.
In other words, we seek the minimum set of physically-meaningful parameters that 
captures the majority of the variance in galaxy properties.

The Fundamental Manifold, so named in reference to its
antecedent, the Fundamental Plane \citep{dd87,d87}, is tightly populated by galaxies
of all morphological types and luminosities.  By asking where star clusters lie relative to this
manifold, we have a means to quantify how distinct star clusters are from galaxies. 
The relationship is straightforward
\begin{equation}
\log r_e = 2\log V - \log I_e - \log \Upsilon_e - C,
\end{equation}
where $r_e$ is the half-light radius in kpc, $V$ represents the internal velocity of the
system and
is a combination of  the velocity dispersion, $\sigma$, and the rotational velocity, $v_r$,
in km s$^{-1}$,  $I_e$ is the mean surface brightness within
$r_e$ in solar luminosities
per square parsec, $\Upsilon_e$ is the total mass-to-light ratio of the galaxy within 
$r_e$ in solar units, and $C$ is a normalization constant.  Additionally,
and what leads to the finding that the systems populate a 2-D manifold, is the empirical
determination that $\Upsilon_e$ is primarily a function of $V$ and $I_e$. The functional
form of $\Upsilon_e(V,I_e)$ is
not simple, hence the broad term manifold, and while the data used to derive the fitted surface extend over many orders of magnitude
along each of the relevant parameters, the function is unconstrained over much of the plausibly accessible parameter space and must be used with caution. For example, the function as described by \cite{zzg} was unconstrained in the region of parameter space occupied by star clusters.

Equation (1) originates
from the virial theorem, and when combined with the finding that $\Upsilon_e = \Upsilon_e(V,I_e)$,
it limits the region of parameter space
occupied by
the galaxy population. In other words, {\sl the virial theorem
alone allows infinitely more possible configurations that those found to populate the FM.} As such,
the FM places a strong constraint on galaxy formation and evolution models. The simplicity of Equation (1) and its relation to the virial theorem satisfies our first condition above. The empirical finding that the scatter about this relation is small for the most diverse set of galaxies available, and
yet captures their variety in kinematics, size, and luminosity, demonstrates that we have included the full dimensionality of the family of galaxies, at least to within the observed scatter, and therefore satisfies our second and third conditions.

Before testing whether star clusters lie on the FM, it is important to understand what
the relationship implies about galaxies. First,
satisfying Equation (1) requires that galaxies be in virial equilibrium and that they have what we will refer to as ``structural
terms"  that have a small degree of
scatter. Hereafter, we use the term structure to encompass both the spatial distribution and kinematics of stars and dark matter. These structural terms are the calculated numerical  
coefficients of the $v^2$ and $GM/r$ terms one would obtain when evaluating the kinetic or
potential energy terms in the virial theorem. For example, an imaginary galaxy of constant
density out to $r_e$ would have a different total potential energy than a galaxy that consists
of the same amount of mass in a ring of radius $r_e$, although the potential energy of either 
galaxy can be expressed as being proportional to $GM/r_e$. It is impossible to numerically tally the 
terms in the virial theorem because one never has complete information regarding the phase space distributions of
the stars and dark matter in a galaxy. Nevertheless, the
low scatter and lack of systematic differences among different galaxy types when applying
Equation (1) indicates that
the structural terms, which we have blithely incorporated into the constant term, are not detectably different among galaxies. Second, and
perhaps somewhat more surprising, is that $\Upsilon_e$ is principally determined by $V$ and $I_e$,
and therefore only modestly related to other factors that might have had an influence on $\Upsilon_e$,
beyond whatever joint influence those factors may have had on $V$ and $I_e$. As such, there is a strong
connection between the overall structure of a galaxy and the mass-to-light ratio within $r_e$. 
It is this second part in particular, the interdependence of $V$, $I_e$ and $\Upsilon_e$, 
that one might
suspect would be qualitatively distinct for star clusters if their formation or evolution mechanism(s) is wholly different than that of galaxies. Consider that dSphs, with their $< 10 {\rm \ km \ s}^{-1}$
velocity dispersions, are quite different in appearance from star clusters with similarly low
velocity dispersion. 

Our plan is to test whether star clusters can be incorporated easily into the FM construct. If they can be, then they can be described
as a low-mass, high-concentration extension of the family of galaxies because the lack of dark
matter within $r_e$ has produced no detectable difference in their structure relative to galaxies.
On the other hand, if clusters lie significantly off the FM then their structure is sufficiently different 
from that of galaxies to suggest differences in their formation and/or evolution. To provide a specific
example, one could imagine a system like a star cluster that is instead dominated by dark matter
(consider turning some fraction of the stars into dark matter). This system would
have the kinematics and size typical of normal star clusters, but a very different surface brightness and mass-to-light ratio. If we then allow the dark matter to follow a different
radial density profile, all the observables would be affected and we would not necessarily expect 
this system to lie on the same relationship between $\Upsilon_e$, $V$, and $I_e$.  The converse does not hold in the sense that systems on the FM do not
necessarily have the same evolutionary history.

Although a better understanding of the origin of galaxies and star clusters is the primary goal, universal scaling relationships provide numerous ancillary benefits. For example, we are familiar with the distance determinations using the Fundamental Plane (FP) and Tully-Fisher (TF) relationships. Unlike those scaling relations, the FM also provides a measure of the total mass-to-light ratio within
$r_e$.  Given that star clusters lack dark matter halos, we will use the FM to 
constrain the {\it stellar} mass-to-light ratio, $\Upsilon_*$, as a function of cluster parameters such
as main sequence turn-off age and color. We will then compare these measures to values calculated using stellar population models as a test of those models.

In summary, by considering star clusters and UCDs, we 
test whether the FM formalism extends to the low-mass, stellar-dominated 
extremum of gravitationally-bound stellar systems in \S3.  
We find that 
star clusters do lie on the FM but with apparently greater scatter than observed among the galaxies. We then explore both observational and physical 
sources of that scatter. We find that there is a correlation between cluster age
and deviation from the FM, in the sense expected for variations in $\Upsilon_e$ due to
the evolution of the stellar population, but in quantitative disagreement with the predictions from stellar population
models. We explore this behavior and present empirical derivations of the relationships between $\Upsilon_e$,
age, and $B-V$ in \S4. We find that the dominant source of scatter in the current data set is
observational, arising from the uncertainties in the cluster velocity dispersions. We use the FM
to make predictions for the velocity dispersions of those clusters without measured dispersions.
In \S5 we summarize our findings and discuss the implications.
In an Appendix we
revisit previous claims that are in apparent conflict with our results \citep{forbes, tollerud} and provide a resolution.

\section{The Data}

We draw the measurements of structural parameters, velocity dispersions, {\sl modeled} stellar population mass-to-light ratios ($\Upsilon_{*,mod}$), ages, colors ($B-V$), and metallicities ($\langle Fe/H \rangle$) from the compilation by \cite{clusters} of 
Local Group star clusters. That sample consists of 153 spatially resolved clusters from a range of nearby galaxies. Within that sample there is a subset of 57 clusters for which they have also compiled velocity dispersion measurements. As such, it includes
clusters of all ages, a wide range of metallicities, and the full range of environments found in the Local Group.

The data for stellar systems ranging from dwarf spheroidals to galaxy cluster spheroids \cite[CSphs to distinguish them from stellar clusters throughout, see][]{zzg} are drawn from \cite{zzg}  and for ultra-compact dwarfs from \cite{mieske}, selecting only 
objects with estimated masses $> 5 \times10^6 M_\odot$ to distinguish the UCDs from the
stellar clusters in their sample.

\section{Star Clusters and the Manifold}

To define the manifold, we use only the spheroidal galaxies and CSphs from \cite{zzg} as our comparison sample to minimize potential
problems arising from composite stellar populations and to limit ourselves to systems supported
by velocity dispersion rather than rotation.
In the vast majority of local
galaxies, either $v_r$ or $\sigma$ dominates, and therefore we have not yet truly tested
whether $V^2 \equiv 0.5v_r^2 + \sigma^2$ is preferable to other parameterizations, such as $V ^2\equiv {\rm max}[0.5v_r^2,\sigma^2]$,
or whether 0.5 is the optimal coefficient for the $v_r$ term \citep[this topic will be addressed in another paper,
but values of 0.3 to 0.5 are plausible in idealized models;][]{weiner06} and such parameterizations
have been used elsewhere \citep{burstein, kassin, covington}.
Here, we avoid questions regarding the incorporation of rotation supported systems in the
FM formalism by only analyzing systems in which $\sigma$ dominates.
The low relative rotation of Milky Way clusters is inferred from their low ellipticities \citep{freeman}. 
This argument is weaker for other cluster populations, such as those in the Magellanic Clouds, where
larger ellipticities are sometimes found. However, even in the Clouds, the cluster ellipticities are generally less than 0.5 \citep{hill}, similar to the ellipticity of one of the most extreme MW clusters, $\omega$ Cen, which has $v/\sigma \sim 0.5$ \citep{omegacen}. Therefore, for the remainder
of this study, we adopt $V\equiv \sigma$.
Using only spheroidal galaxies further simplifies our work by removing systems with ongoing or recent
star formation, for which $\Upsilon_e$ may have a dramatic dependence on mean stellar age.

\begin{figure*}
\epsscale{0.8}
\plotone{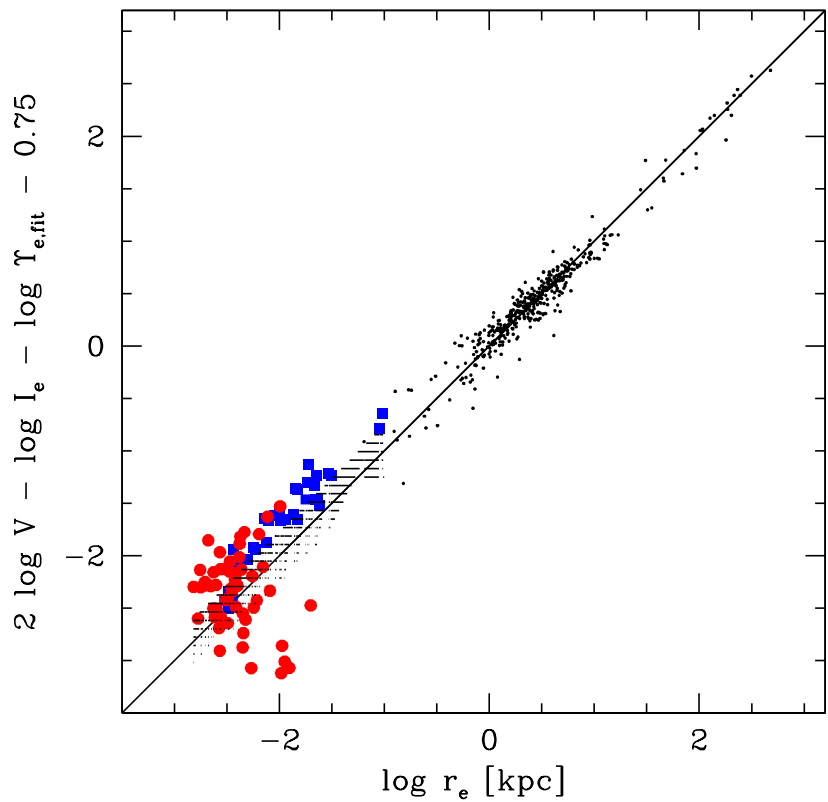}
\caption{The edge-on view of the FM for spheroidal stellar systems, including star clusters and ultra-compact dwarf galaxies (UCDs).  We are extrapolating the fit from the region populated
by systems with $\log r_e > -1$ (i.e. excluding both UCDs and stellar clusters).
We illustrate the statistical uncertainties in that extrapolation 
with dots representing projections of the surface where individual fit coefficients are altered by
$1\sigma$. 
Spheroidal stellar systems from \cite{zzg} are
plotted as filled black circles, clusters with tabulated $\sigma$'s \citep[][and references therein]{clusters}
are plotted as filled red circles, and UCDs with $M > 5\times 10^5 M_\odot$ from \cite{mieske} are plotted as filled blue squares. The extrapolated fit, within its $1\sigma$ uncertainties, passes through the majority of the 
stellar clusters and UCDs, although the scatter among the clusters is greater than among the galaxies. Uncertainties are not plotted, but the dominant uncertainty in the cluster population is discussed in \S\ref{sigsig} and the uncertainties in 
the comparison galaxy population are comparable to the scatter \citep{zzg}.The clusters that deviate well below the fit are the youngest clusters.}
\label{fig:fit}
\end{figure*}

The final relevant technical issue with the galaxy sample is the heterogeneous nature of the source samples and how one should compare it to the cluster sample. In particular,
the cluster sample contains $V$-band photometry, while the galaxy sample contains a range
of photometry from $V$ to $I$-bands. Equation (1) is band-independent in that $I_e$ and $\Upsilon_e$
for a particular object need to be in the same band, but that band can be different for different objects.
However, the manifold that we fit to describe $\Upsilon_e$, $\Upsilon_{e,fit}$, is band-specific and has been 
derived using the heterogeneous galaxy sample, corrected to the $I$-band. The values of $I_e$ used to determine $\Upsilon_{e,fit}$ are in solar units, so the filter dependence arises from the difference in colors between the galaxy and the Sun when converting $L_e$ to $I_e$. In effect we are assuming solar colors for the galaxies,  although any uniform color difference among all galaxies is absorbed into the calibration of $C$ in Equation (1). The remaining concerns are color differences that vary across the sample and band-related differences in $r_e$. If these
exist, then they are inappropriately incorporated into our fit for $\Upsilon_e$.
One clear path to progress is
obtaining a single, uniform data set for the further study of the FM. However, in the interim, we proceed with the 
motley sample available and caution that for systems with colors much different than solar
colors there will be a yet unknown color-correction to $\Upsilon_{e,fit}$. The empirically-confirmed small scatter among galaxies, even when including star forming galaxies \citep{zzg}, suggests that this effect will be negligible for a sample consisting only of spheroidal galaxies, but could be a factor among the star clusters, which span a wider age range.

\begin{figure}
\epsscale{0.8}
\plotone{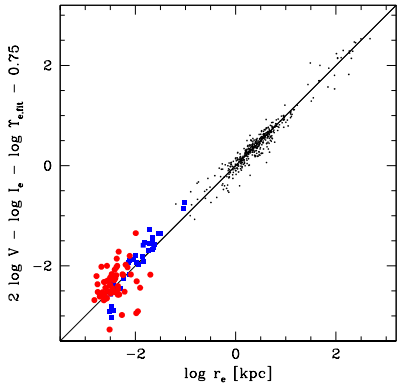}
\caption{The edge-on view of the FM for spheroidal stellar systems, where the empirical fit
for $\Upsilon_{e,fit}$ includes both the stellar clusters and UCDs. Symbols are as described for
Figure \ref{fig:fit}.}
\label{fig:fit1}
\end{figure}

We revaluate $\Upsilon_{e,fit}$ using only those galaxies, the spheroids, from \cite{zzg}
that are $\sigma$ dominated, including galaxy cluster spheroids (CSph).
In doing so, we have also reevaluated the constant $C$ in Equation (1) using the \cite{cappellari} sample, for which they have full dynamical modeling fit to 2-dimensional spectral data for
spheroidal galaxies, and find $C = 0.75$. 
In Figure \ref{fig:fit} we show where the 57 globular clusters with measured velocity dispersions
and the UCDs with $M > 5\times 10^6 M_\odot$ from \cite{mieske}
fall when using this fitting function, {\it which was not derived using either the star
clusters  or UCDs}. 
In addition, we show with small points the range of possible locations in this projection 
of the extrapolated relationship covered by random, $\le 1 \sigma$ deviations of each of the fitted coefficients. The star clusters and UCDs lie within the extrapolated 
version of $\Upsilon_{e,fit}$ given the fitting uncertainties.
Because the extrapolated fitting function works well in the mean for stellar clusters, we infer that clusters are indeed the low-mass, limiting case of high-stellar-mass-fraction systems and refit
$\Upsilon_{e,fit}$ using the cluster and UCD data (Figure \ref{fig:fit1}). The resulting fit is 

\begin{eqnarray}
\Upsilon_{e,fit} &  = & 1.49 - 0.32 \log V - 0.83 \log I_e \nonumber \\
 & & {} + 0.24\log^2 V +0.12 \log^2 I_e  \nonumber\\
 & & {} - 0.02 \log V \log I_e 
\end{eqnarray}
and is used hereafter.

\begin{figure}
\epsscale{0.8}
\plotone{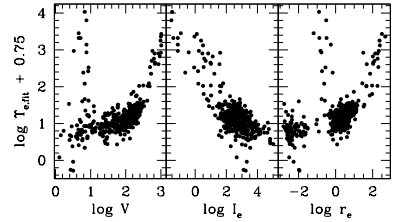}
\caption{Projection of $\Upsilon_{e,fit}$ along the three observable parameters. The plume of
objects rising in the left of the first and third panels are the dSphs. The objects near the bottom
left of the first and third panels are the star clusters and UCDs. It is difficult in projection to determine
whether a single surface fits all the data and, if not, which population to fit. We find that a single surface does fit all the objects.
}
\label{fig:proj}
\end{figure}

Projections of this function along the various axes (Figure \ref{fig:proj}) illustrate how difficult it is to see the multidimensional nature of the distribution and how certain populations of objects appear to stand out as different despite populating the same surface. Because there is no physical motivation for this particular form, the addition of new data, particularly those that probe a new part of parameter space, could require changing the functional form. However, the implications of the FM are  independent of the particular functional form, depending only on the finding that $\Upsilon_{e,fit}$ is principally a function of $V$ and $I_e$. The apportionment of the constant term between $\Upsilon_{e,fit}$ and the $C$ in  
Equation (1) is set by the external calibration to independently measured values of $\Upsilon_e$ and by construction in  Equation (1) that sets the two sides to be equal.
The scatter about the FM is evidently larger for the star clusters than it is for the galaxies \citep[0.35 vs. $\sim$ 0.1;][]{zzg}, so we turn to discussing possible sources of that scatter.

There are various potential causes of the larger scatter, 
including structural variations among clusters that are not present in galaxies,
proportionally larger uncertainties in at least one of the observed parameters for the clusters, and the extrapolation of a fitted function beyond the parameter range over which it is constrained. Among the observed parameters, the likely source of scatter is $V$, which is more poorly measured in these low-velocity systems than in normal galaxies, and so 
we will begin our discussion by assessing the impact of the quoted uncertainties in $V$. 
However, of greater interest is the possibility that some of the scatter is driven by a
parameter that has not been included in our expression for $\Upsilon_{e,fit}$. 
This type of effect is often refereed to
as a ``second parameter" question in correlations, although in our case it would be a third
parameter. If there is such a parameter, it has been inadvertently neglected in our formulation of
the fitting function for $\Upsilon_{e,fit}$. As such, $\Upsilon_{e}$ as evaluated using Equation (1), $\Upsilon_{e,FM}$, should correlate with this yet unidentified parameter (as it does with $V$ and $I_e$). 

 In closing, we note that there are two studies \citep{forbes, tollerud} that
appear to conflict with our claim that stellar clusters satisfy galaxy scaling relations. 
We describe our interpretation
of those studies in Appendix A and conclude that neither study is truly in conflict with the results
presented here.

\subsection{Observational Source of Apparent Scatter: Velocity Dispersions
\label{sec:sigma}}

\begin{figure}
\epsscale{0.8}
\plotone{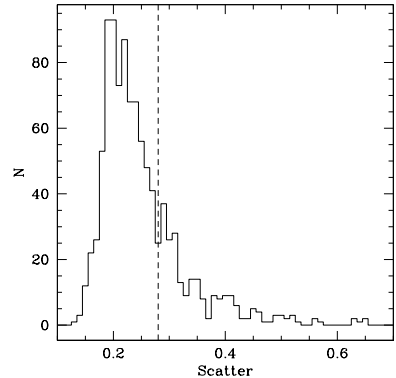}
\caption{Distribution of simulated FM scatter for star clusters. We assign each cluster a value of $\sigma$ that places it exactly on the FM. We then use the quoted uncertainty in $\sigma$ to create random samples. The plot shows the resulting scatter distribution for 1000 samples. The dashed line shows the measured scatter. Although the observational scatter is larger than the typical
simulated one, we cannot confidently exclude the possibility that it is only the result of the fractionally larger uncertainties in $\sigma_{OBS}$ for these low $\sigma$ systems.}
\label{fig:scatter}
\end{figure}

For these low velocity dispersion systems, the uncertainties in $\sigma$ are fractionally
the largest of all the input parameters. We now explore the level of scatter about the
FM expected simply due to the quoted uncertainties. 
We use the 51 old ($\log({\rm age}) > 10$ Gyr) clusters with published velocity
dispersion measurements and associated uncertainties  to avoid any possible
effect of age on the scatter.
We place each cluster exactly on the FM by adopting a value for $\Upsilon_e$ that
satisfies Equation (1). We then draw
a thousand cluster samples using 
Gaussian distributed errors on the velocity dispersion, using the corresponding value for the uncertainty in $\sigma$ 
for each particular cluster. In Figure \ref{fig:scatter} we
plot the distribution of the simulated scatter for the thousand trials and the observed scatter for comparison.
The observed scatter is somewhat larger than the typical simulated one, but we cannot
exclude the possibility that the observed scatter is directly due to the $\sigma$ uncertainties with greater than 80\% confidence. We conclude that the observed scatter in Figure \ref{fig:fit} is
principally due to the uncertainties in $V$, but
cannot conclude that no physical sources
of scatter remain. We test for physical sources of scatter in the next three subsections.

\subsection{Physical Sources of Apparent Scatter: Age}
\label{sigsig}

\begin{figure}
\epsscale{0.8}
\plotone{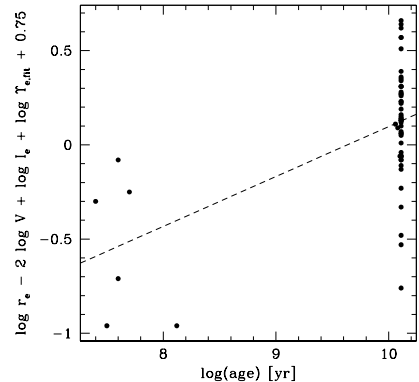}
\caption{Deviations from the FM vs. age for the clusters in \cite{clusters} sample.
The dashed line represents the best fit line to the relation.
As expected, the mean deviation is nearly zero for older systems, whose populations are most like those of early-type galaxies, from which we derived the functional form of $\Upsilon_{e,fit}$. The scatter is larger separately for either the old and young populations than for the galaxies, but that is mostly, if not wholly, due to the clusters' proportionally larger $\sigma$ uncertainties (see \S\ref{sec:sigma}).}
\label{fig:fm-age}
\end{figure}

There is at least one physical cause of scatter that must play a role, that we can quantify easily, and that
we can model --- variations in the stellar population mass-to-light ratio, $\Upsilon_*$.
In Figure \ref{fig:fm-age} we demonstrate that the observed deviations from the FM for star clusters are, as expected,
different for the young and old stellar systems. We fit a linear relation in
$\log ({\rm age})$ to describe the residual ($0.272(\log({\rm age}) - 10.11) + 0.044)$ and  subsequently use this fit and the cluster ages from \cite{clusters}, obtained from resolved stellar main sequences, to 
correct $\Upsilon_{e,fit}$ for this systematic age deviation. Importantly, the residual is small for the older clusters, for which
the calibration to galaxies is expected to be most appropriate.
There remain, however, at least two open questions. First, the scatter remains large
for both young or old clusters (although, at least for the old clusters, we demonstrated above
that observational uncertainties can account for the scatter). Second, the mean trend only testifies to some age-related effect,
not that this effect must be related to $\Upsilon_*$.
There are three reasons that the trend is likely due to the evolving stellar population rather than to age-dependent dynamical effects: 1) the stellar population must, to some degree, evolve and affect $\Upsilon_e$ because $\Upsilon_e$ is related to $\Upsilon_*$, 2) the fitting function works best (i.e. the residual is $\sim$ zero) for the oldest clusters, for which the calibration for spheroidal galaxies and their similar stellar populations should be most appropriate, and 3) dynamical effects, such as those involving two-body relaxation, should trend in the opposite direction with age, becoming more, rather than less, important in the older clusters. Unfortunately, further discussion is limited by the lack of clusters with measured $\sigma$ across our entire age range, 
$8 < \log ({\rm age [yrs]}) < 10$. Although the entire \citet{clusters} cluster sample does contain clusters in this age range, 
we need an estimate of $\sigma$  to include them in our discussion. We next
describe how we use the FM to obtain such estimates.

\begin{figure}
\epsscale{0.8}
\plotone{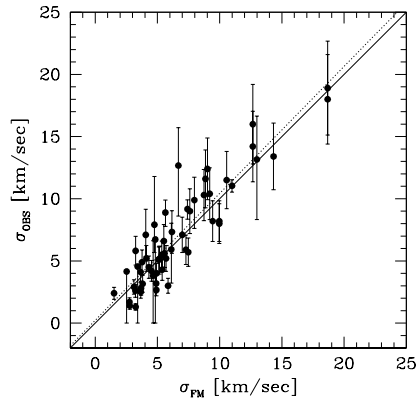}
\caption{Comparison of $\sigma$ estimated from 
placing clusters on the FM ($\sigma_{FM}$) to their measured velocity dispersions ($\sigma_{OBS}$).
The solid line represents the 1:1 line, while the dashed line represents the best fit line. The procedure works well, so we apply it to the entire \cite{clusters} sample to estimate $\sigma$ for
those clusters that do not yet have $\sigma_{OBS}$.}
\label{fig:sigtest}
\end{figure}

Because the right hand side of Equation (1) can be 
written as only a function of $V$ and $I_e$ after adopting $\Upsilon_{e,fit}$,
we use the tabulated values of $I_e$ and $r_e$ for the clusters to
solve for $V$, which we equate with $\sigma$ (see \S3). 
The drawback of this approach is that we have now assumed, on the basis of Figure \ref{fig:fit}, that both Equation (1) and
$\Upsilon_{e,fit}$ are applicable to clusters.
We account for the trend of $\Upsilon_{e,fit}$ with age shown in Figure \ref{fig:fm-age} using the linear fit, and then we
solve the equation for $V$.  
We test this procedure on the subset of 57 clusters with measured $\sigma$ in
Figure \ref{fig:sigtest}. The agreement is encouraging and we proceed to 
estimate $\sigma$ for all 153 clusters.  
The agreement within the uncertainties of the calculated and observed $\sigma$'s in Figure \ref{fig:sigtest}  demonstrates that much of the scatter among clusters
on the FM must come from the uncertainties in their respective $\sigma$'s, as we had surmised on other grounds.  In other words, we
have not created a situation where we absorb errors in other quantities or intrinsic scatter
into our derived values of $\sigma$.  
In all subsequent plots, we use the inferred $\sigma$'s, but highlight those clusters with measured $\sigma$'s using large open circles to illustrate the degree to which any trends are driven by
those clusters with unconfirmed velocity dispersion estimates.

An interesting subset of clusters are those of lower surface brightness in the outskirts of the
Milky Way. Because of their distance and naturally low
$\sigma$, they tend to not have $\sigma$ measurements of high relative precision.
Outside of
the compilation of cluster properties by \cite{clusters} that we have used exclusively, there are measurements of $\sigma$
for Pal 5 \citep{dehnen} and Pal 14 \citep{jordi}. Using these values, we find that 
even these clusters fall on the FM relation.
As a target for future observers, we present the calculated values of $\sigma$
necessary to place similarly unusual clusters on the FM (Table \ref{tb:oddballs}). These low dispersions illustrate why high precision
spectroscopy and a detailed knowledge of the binary populations (cf. \cite{dehnen}) 
are needed to confirm the calculated $\Upsilon_e$'s and to resolve the question of whether such systems lie on
the FM.

\begin{figure}
\epsscale{0.8}
\plotone{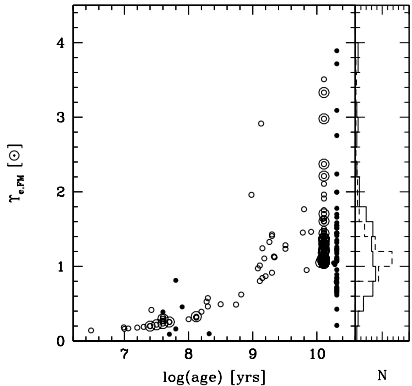}
\caption{Total mass-to-light ratio within $r_e$ as evaluated using Equation (1), $\Upsilon_{e,FM}$, vs. main sequence turn-off age compiled by
\cite{clusters} using stellar color-magnitude diagrams from a set of references. All clusters are plotted using the estimated $\sigma$'s, $\sigma_{FM}$ (small open circles). Those clusters that also have a measured $\sigma$, $\sigma_{OBS}$, are highlighted (large open circle). For comparison, we recalculate $\Upsilon_{e,FM}$ using 
$\sigma_{OBS}$ when available (filled circles that are
offset in age by 0.2 for clarity). 
On the right, we show the distribution of $\Upsilon_{e,FM}$ for the old ($\log(\rm{age}) >$10) clusters. The dashed line represents the distribution  using $\sigma_{FM}$, while the solid line represents the distribution using  $\sigma_{OBS}$ (the histograms are scaled arbitrarily for better comparison). The inclusion of all clusters supports the conclusion from Figure \ref{fig:fm-age} that there is a correlation of $\Upsilon_{e}$ with age.}
\label{fig:allages}
\end{figure}

We now return to investigate the inferred drift in $\Upsilon_e$ vs. age by including
all star clusters with or without measured $\sigma$'s. We plot the total mass-to-light ratio within $r_e$ evaluated using Equation (1),  $\Upsilon_{e,FM}$,
vs. age (Figure \ref{fig:allages}). For those clusters where we estimate
$\sigma$, the values of $\Upsilon_{e}$ are not independent of the measured age, and
so trends should be viewed with caution and compared to those obtained from clusters
with measured $\sigma$.
Because stellar clusters are dark matter free, $\Upsilon_e = \Upsilon_*$. 
We confirm that the value of $\Upsilon_e$ necessary to place clusters on the FM
is a simple function of age and, as seen in Figure \ref{fig:allages}, that the sense of the trend (higher values
of $\Upsilon_e$ for older ages) is as expected and matches the behavior
determined from only the clusters with measured $\sigma$. 
Because all but the lowest four
of the clusters in Figure \ref{fig:fit} are old, the
apparent scatter is not the result of age differences within the sample.

\subsection{Physical Sources of Apparent Scatter: Structural Terms}

\begin{figure}
\epsscale{0.8}
\plotone{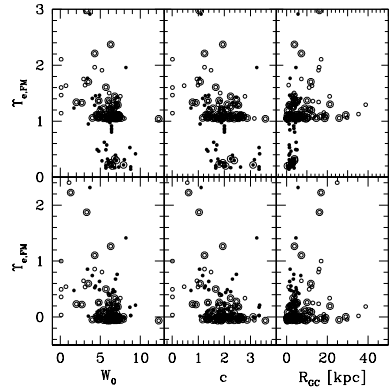}
\caption{Examination of structural effects on $\Upsilon_{e,FM}$.
We plot $\Upsilon_{e,FM}$ using $\sigma_{FM}$ for the entire cluster sample vs. various structural measures.
In the upper panels, we plot $\Upsilon_{e,FM}$ uncorrected for the age relationship from Figure \ref{fig:fm-age}, and in the lower panels we correct for the age relationship. 
Clusters older than 10 Gyr are represented by small open symbols and younger ones by filled symbols. The larger open circles again highlight those clusters with measured $\sigma$'s, although
we use $\sigma_{FM}$ for all clusters. 
We plot $\Upsilon_{e,FM}$ vs. a dimensionless measure of the central potential, $W_0$ (leftmost panels),  vs. concentration, $c \equiv \log(r_{tidal}/r_{core})$ (center panels), and 
vs. the distance of each cluster to the center of its parent galaxy (rightmost panels).
That values of $W_0$ and $c$ come from \cite{clusters} for their preferred Wilson models. The magnitude of 
any correlations 
are significantly smaller than
the age effects visible between the lower and upper panels. A small number of clusters ($\le 5$)
lie beyond the plotting limits, which are chosen to bring out details of the distribution of the bulk of the 
clusters.}
\label{fig:comp_dyn}
\end{figure}

Another potential source of scatter is variance related to the cluster structural properties. Star clusters 
can evolve dynamically due to their relatively short dynamical times (in comparison to 
the dynamical times for galaxies) and therefore should have larger variations in
their structural terms, which reflect both the stellar kinematics and spatial distribution. 
We examine whether structural differences 
among the clusters have been incorrectly excluded from Equation (1), or $\Upsilon_{e,fit}$, by looking for correlations between $\Upsilon_{e,FM}$ and measurements 
of the structural state of the cluster. In a similar approach,
\cite{mclaughlin} found some evidence for correlations between the structural state of Galactic clusters and their location relative to the FP. Figure \ref{fig:comp_dyn} shows
the relationships between  $\Upsilon_{e,FM}$ and central potential depth, concentration, and galactocentric distance. Whatever trends there may be in the lower panels, they are clearly of much lower magnitude than that with age, which is evident as the bimodal distribution in the upper panels of the Figure. We
conclude that none of the measures of cluster structure that are 
currently available suggest a relationship between deviations from the FM and the dynamical state of the cluster beyond any directly related to age. Because dynamical state depends on more than just age, we conclude that stellar population differences, which do depend only on age, are the primary driver of the trends we observe. 

\begin{figure}
\epsscale{0.8}
\plotone{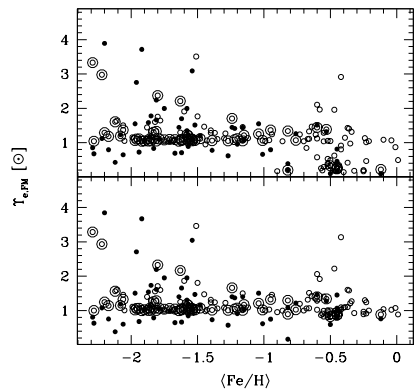}
\caption{The dependence of  $\Upsilon_{e,FM}$ on cluster metallicity compiled by
\cite{clusters} from a set of references. $\Upsilon_{e,FM}$ is evaluated using $\sigma_{FM}$
for all clusters (small open circles). Clusters for which $\sigma_{OBS}$ is available are
highlighted with large open circles. We recalculate $\Upsilon_{e,FM}$ using $\sigma_{OBS}$ and plot those results for comparison as filled circles. 
In the bottom panel we correct for the age dependence of $\Upsilon_{e,FM}$ inferred from Figure \ref{fig:fm-age}. 
The age-related effects are significantly larger than any correlation with metallicity.
}
\label{fig:fe-corr}
\end{figure}

\subsection{Physical Sources of Apparent Scatter: Metallicity}

Next, we search for a dependence of $\Upsilon_{e,FM}$
on cluster metallicity (Figure \ref{fig:fe-corr}).
There is no evident dependence of $\Upsilon_{e,FM}$ on metallicity below $\langle Fe/H\rangle < -1$ in the upper panel of the Figure. There may be an effect for higher metallicities, although 
that is where age differences become important as well.  In the lower panel, we apply the 
age correction derived from Figure \ref{fig:fm-age} and find that there is little if any residual metallicity dependence. Again, the principal driver of the trends appears to be age. 

\subsection{Physical Sources of Apparent Scatter: Conclusions}

On the basis of the correlation between scatter and age, and the agreement  between the
inferred sense of the effect and the qualitative expectations from stellar population models, we
conclude that age is the dominant physical source of scatter. However, our data do not
allow us to eliminate all possible structural differences that might play a role, nor do
they statistically demonstrate the superiority of age as a driving parameter over
the other parameters now available.
The peculiarities of the sample, only six clusters with measured $\sigma$ at age $\sim 10^8$ yrs and the remainder all at old ages,  preclude application of statistical tests such as 
principal component analysis or partial correlation coefficients as a means to identify a dominant physical effect. However, this shortcoming can be remedied with reasonable observational effort.

\section{Estimating $\Upsilon_*$}

\begin{figure}
\epsscale{0.8}
\plotone{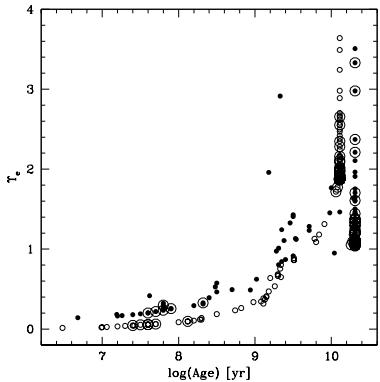}
\caption{Direct comparison of two ways to evaluate $\Upsilon_e$ as a function of age. We plot the results using the stellar mass-to-light ratio from stellar population models, $\Upsilon_{*,mod}$ (small open circles),  and from the FM using Equation (1) and $\sigma_{FM}$, $\Upsilon_{e,FM}$ (filled
small circles). Large open circles highlight clusters with $\sigma_{OBS}$. The ages
for the values of $\Upsilon_{e,FM}$ have been arbitrarily shifted by 0.2 for clarity.
The two methods of estimating $\Upsilon_e$ qualitatively track each other with age, 
but differ in the overall amplitude of the change from the youngest to oldest clusters.}
\label{fig:comp}
\end{figure}

\begin{figure}
\plotone{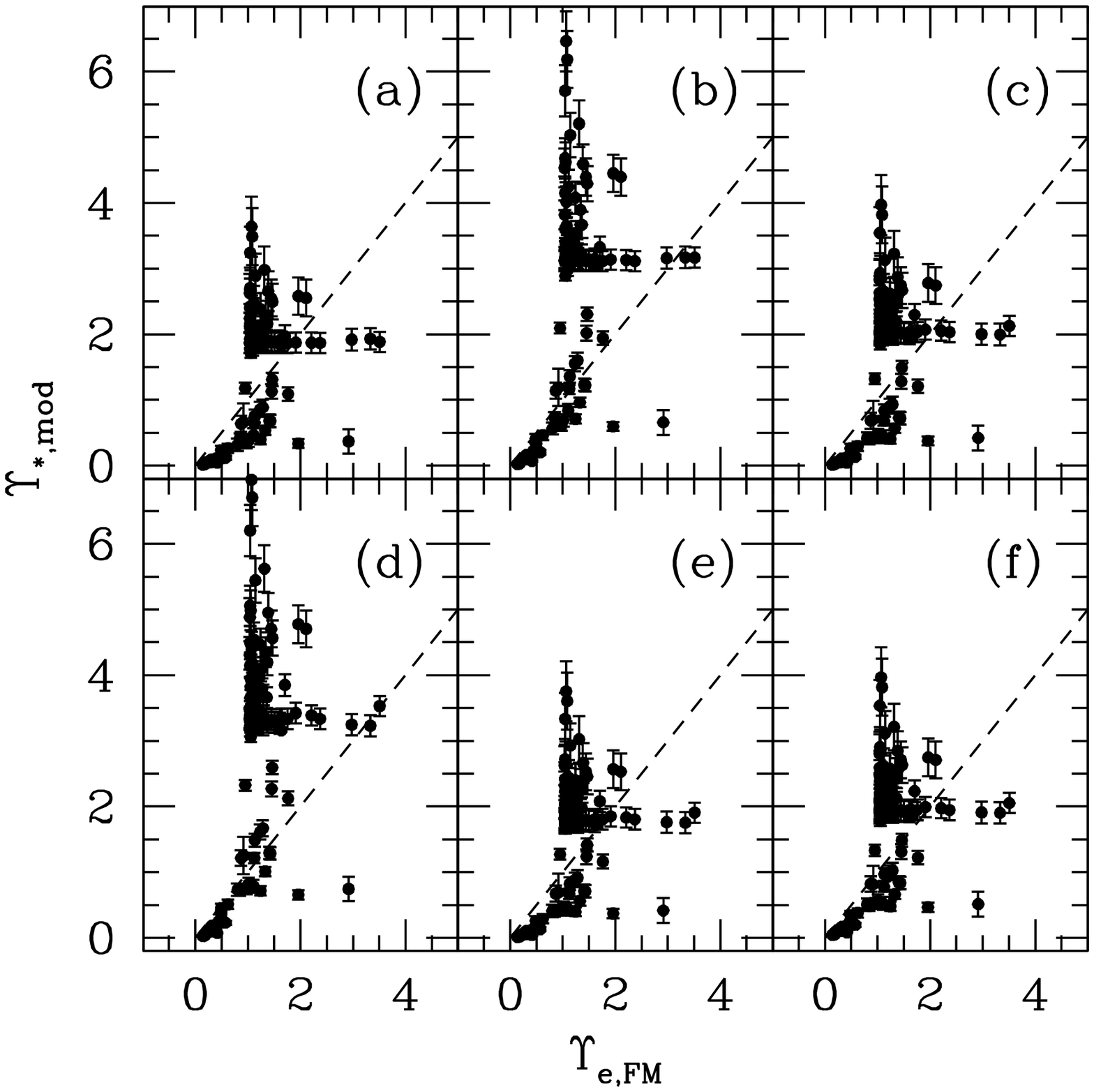}
\caption{Comparison of $\Upsilon_{*,mod}$ and $\Upsilon_{e,FM}$ for different stellar population models. Panels represent (a) Bruzual-Charlot models \citep{bc} with Chabrier IMF \citep{chab}; (b) Bruzual-Charlot models with Salpeter IMF; (c) P\'EGASE models \citep{pegase} with Chabrier IMF; (d) P\'EGASE models with Salpeter IMF; (e) P\'EGASE models with the Chabrier globular cluster IMF \citep{chab}; and (f) P\'EGASE models with the \cite{kroupa} IMF. The sense of the deviation for the oldest ages (highest values of $\Upsilon_e$) is consistent among all varieties of stellar models. The stellar models consistently over predict $\Upsilon_e$ relative to the FM estimates for these older clusters.}
\label{fig:stpops}
\end{figure}

\subsection{Comparing $\Upsilon_{*,FM}$ to Stellar Population Models}

The variation in $\Upsilon_{e}$, which for stellar clusters is equivalent to $\Upsilon_*$,  vs. age has a natural explanation in terms
of stellar population evolution.  In Figure \ref{fig:comp} we compare estimates of 
$\Upsilon_e$ based on the FM, $\Upsilon_{e,FM}$, to those based
on the model values of $\Upsilon_*$, $\Upsilon_{*,mod}$, tabulated by \cite{clusters} using their preferred Bruzual-Charlot models with a Chabrier initial mass function (IMF).  
The sense of the behavior of $\Upsilon_e$
with time is consistent for both $\Upsilon_{e,FM}$ and $\Upsilon_{*,mod}$ suggesting that we are indeed
seeing the effects of stellar evolution on $\Upsilon_e$. However, the quantitative
agreement is poor, with $\Upsilon_{e,FM}$ overestimating $\Upsilon_{*,mod}$ at young ages
and underestimating it at old ages. 

The existence of a significant difference between $\Upsilon_{*,mod}$ and $\Upsilon_{e,FM}$, particularly at old ages, is independent of the particular model or initial mass function, among those currently in general use. We compare $\Upsilon_{e,FM}$ to the full range of $\Upsilon_{*,mod}$ presented by \cite{clusters} in Figure \ref{fig:stpops}. We find similar results after comparing to  2007 Bruzual-Charlot models, which are not included in \cite{clusters}. The 
magnitude of the differences between $\Upsilon_{*,mod}$ and
$\Upsilon_{e,FM}$ is similar to that of the differences among the stellar population models themselves. It is
therefore possible that there are problems with the stellar population models at this
level. \cite{conroy} explore the uncertainty in stellar population models, the
bulk of which comes from modeling intermediate and low-mass stars near the end of their lives where they
become far more luminous but for short periods of time. \cite{tonini} show that including the thermally-pulsating AGB phase can increase the emission from an intermediate age galaxy by 1 magnitude in the $K$-band. The level of the discrepancy suggested by our FM analysis is consistent with these potential problems in 
$\Upsilon_{*,mod}$. Even at the young (low $\Upsilon_e$) end of things, there 
are clear differences among the models, with those using the Salpeter IMF producing results more
in line with those from the FM. 
Further discussion of the details of stellar population models
is beyond the scope of this paper, but the FM provides a new constraint.
The FM-derived values of $\Upsilon_*$ suggest
that all currently popular stellar population model variants over-predict the stellar mass-to-light ratio of old populations \citep[see][for an alternate approach that results in the same conclusion]{mieske}.

Stellar population models that provide accurate stellar mass-to-light ratios are manifestly important for a variety of uses. However, star clusters on the FM provide a new way to estimate $\Upsilon_e$ for large sets of {\it galaxies}. Currently, the FM zero point ($C$ in Equation (1))
is calibrated using the results for $\Upsilon_e$ from detailed dynamical studies of several tens of early-type galaxies \citep{cappellari}. If we knew, from stellar models, the values of $\Upsilon_*$,
and hence $\Upsilon_e$, for star clusters, we could use them instead to calibrate the FM. 
Once the FM is accurately calibrated, one can use it to solve for $\Upsilon_e$ for any galaxy with measured $V$, $r_e$, and $I_e$. 

\subsection{An Independent Prediction of $\Upsilon_*$}

\begin{figure}
\epsscale{0.8}
\plotone{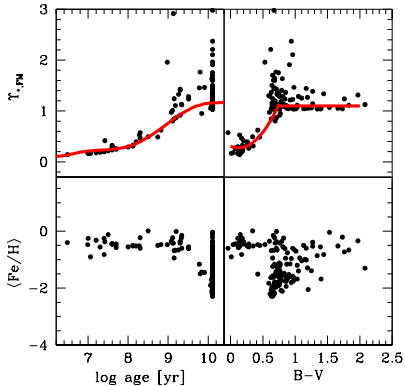}
\caption{We equate $\Upsilon_{e,FM}$ with $\Upsilon_*$ by assuming that star clusters are dark-matter free. In the upper panels we plot the inferred value of $\Upsilon_*$, $\Upsilon_{*,FM}$, vs. age and color. The  curves are the by-eye fits to the data, with coefficients given in Tables \ref{tb:fitage} and \ref{tb:fitbmv}. The fit vs. color is valid for $B-V < 0.75$ and the data imply a constant value of $\Upsilon_*$, 1.13, for colors redder than 0.75. The bottom panels show the trends vs. metallicity. We can see what types of populations are not constrained by the current data (e.g., young, metal-poor populations) and for which the fitting functions fail (e.g., old, metal-rich clusters). }
\label{fig:independent}
\end{figure}

We apply the FM to derive empirical relationships between parameters
such as main sequence turn-off age and color,  and $\Upsilon_*$. Figure \ref{fig:independent}
shows our crude fits
to the age-$\Upsilon_*$ (Table \ref{tb:fitage}) and (B-V)-$\Upsilon_*$ (Table \ref{tb:fitbmv};  the fit is valid for $B-V \le 0.75, {\rm \  for\ } B-V > 0.75 {\rm \ use\ } \Upsilon_* = 1.13$) relationships. Neither 1-D relationship appears satisfactory for the full range of clusters, and certain populations, such as young and metal-poor populations, are completely unconstrained. More sophisticated and well-constrained fits will be possible once more data become available for intermediate age clusters, providing an
alternative to $\Upsilon_*$ estimates that depend on stellar populations models. 

\section{Summary}

We find the following:

1) On average, star clusters fall along the extrapolation of the Fundamental Manifold (FM) defined by spheroidal galaxies. Their larger scatter is  consistent with the proportionally larger uncertainties in their velocity dispersion measurements. 
Even extreme clusters such as Pal 5 and 14 fall on the relationship
when precise velocity dispersions are available.

2) Individual clusters are offset from the FM in proportion to 
their age:  older clusters fall on the manifold and younger clusters do not.
Unfortunately, the sample of clusters with measured velocity dispersions does not include many clusters
with $\log ({\rm age\ [yr]}) < 10$, so we apply the FM relationship to infer $\sigma$ and $\Upsilon_e$ simultaneously for the remainder. This procedure works faithfully for those with
measured $\sigma$. 

3) Aside from the trend with age, deviations from the FM do not correlate measurably with metallicity, dynamical state, or galactocentric distance. 

4) The sense of the evolution of mass-to-light ratio with age is as predicted by stellar population synthesis models, but currently popular models fail to reproduce 
the estimates from the FM \citep[see also][]{mieske}. Specifically, the models over-predict the mass-to-light ratio
of old, metal-rich populations. We provide empirical formulae with which to estimate the stellar mass-to-light ratios of certain cluster populations derived from the FM-estimated mass-to-light ratios.

All stellar systems that we have examined in detail so far--- brightest cluster galaxies \citep{zgz},
various  galaxy types and luminosity classes \citep{zzg, mieske}, and now star clusters --- satisfy a simple, 3-parameter (2-dimensional, but not planar)
scaling relation. For star clusters, deviations from 
the scaling law are age dependent, which is qualitatively consistent with the expectation for
passive stellar evolution of the population. Quantitatively, however, this description fails. We 
suggest, but have not definitively demonstrated, that this disagreement arises from systematic
errors in the stellar population models that lead to an overestimation of the V-band mass-to-light
ratio of old stellar populations. The sense and magnitude of the effect are consistent with
conclusions reached by others investigating stellar evolution models in detail \citep{tonini, maraston}
and at the very least suggest that a factor of two uncertainty remains in modeling stellar populations.
As stressed by others \citep[cf.][]{maraston,conroy}, this uncertainty propagates in 
complicated ways, particularly for composite populations of unknown age distributions, in 
analyses of galaxy evolution. 

We conclude that there are no evident structural differences, in either stellar spatial distribution or kinematics, between galaxies and star clusters beyond those of scale and those captured in 
the mass-to-light ratios within $r_e$. Star clusters appear to be baryon-dominated versions of highly-compact, low-mass galaxies. As such, we are not surprised by the difficulties encountered by others in distinguishing clusters from ultra-compact galaxies.
If there are wholesale differences between the formation process and/or evolution of star clusters and galaxies, they do not manifest themselves directly in the gross physical properties. We conclude that even if significant differences exist between the populations, due to the existence or lack of a dark matter halo, simple measurements will not reveal them. Projections of parameter space in which these populations
are separate are deceptive and should not be taken to indicate that these objects are discrete
families of gravitationally bound stellar systems.

Although this lack of evident distinction is a set-back for attempts to differentiate the populations and probe potential differences in their formation and evolution, it is an advantage to those efforts to build a simple set of models for galaxies. Star clusters can provide an anchor for models of galaxies, empirically defining the stellar mass-to-light ratios or, alternatively, the zero point of the FM. The latter would enable us to use the FM to determine the 
mass-to-light ratio within $r_e$, to a level of precision comparable to that obtained with detailed dynamical modeling or strong gravitational lensing modeling,
for any galaxy with known $V, r_e$, and $I_e$.

 We return to the question of the defining characteristics of a galaxy.
Among the stellar systems that we study, we have not identified a
criterion that we consider a discriminating feature.
Despite the popularity of the dark matter halo criterion,
there exist ``galaxies" that may 
not have a dark matter halo \citep[tidal dwarf galaxies;][]{barnes} or, at least, 
a dominant dark matter halo
\citep[ultra-compact dwarf galaxies;][]{mieske}. In fact, from state-of-the-art optical data alone,
it is difficult to demonstrate that even giant ellipticals contain dark matter \citep{cappellari06}.
An alternative, and more easily observable, classification is the
relative concentration of luminous baryons to dark matter. In such a scheme, systems
composed solely of baryons are not distinct from those with dark halos and whose observed kinematics
are dominated by baryons.  While this scheme makes it harder to connect observations with dark matter simulations,
the reality is that, for systems whose baryons are extremely concentrated, we 
cannot determine whether a halo exists or what its properties might be.

Our primary empirical result is that the existing description of the Fundamental Manifold \citep{zzg} is valid
over an even larger parameter range than previously demonstrated and that
now includes stellar clusters. The existence of a FM for the entire family of stellar systems implies
1) that the objects are virialized and 2) that there is limited variation in the overall
structure. While the former is unsurprising, the latter
implicitly places strong limits on how the systems form and how the luminous baryons settle. 
By extending the validity of the FM, or any scaling relation, over a larger range of parameter
space and a wider set of objects, we are demonstrating that the physics that 
constrains the resulting baryon and dark matter distributions in stellar systems is more
general than previously thought. In other words, whereas one might not have been 
surprised by the implication of the Fundamental Plane that all ellipticals
have related structural properties, it is certainly a puzzle to find that 
spirals, dwarf spheroidals, and now stellar clusters do too. 

While the structure of one stellar system is not a simple scaling of another, there exists a 
transform function that relates the properties of stellar systems. Specifically, we have 
presented a
fitting function for $\Upsilon_e$ that is valid across all stellar systems. 
The generality of this transform function
implies that 1)  the structure of one type of stellar system is not a distinct result of a process particular to that
set of systems, such as violent relaxation for elliptical galaxies, but is instead
the result of an interplay of all processes responsible for 
the generic settling of baryons in gravitational potential wells, 2) the existence of the transform function for
$\Upsilon_e$, and hence the nature of baryon settling, is independent of whether the
system is embedded within a dark matter halo, and 3) differences in initial conditions or
stochastic events during the lifetimes of stellar systems do not invalidate the transform function. 
By including stellar clusters, which we presume are devoid of dark matter, we demonstrate 
that the nature of the transform function for $\Upsilon_e$ depends more on the baryon physics
than the source of the gravitational potential. 
The presence or absence of dark matter imprints no clear feature in
the structure or kinematics of stellar clusters versus galaxies. 
By including stellar clusters, which we presume formed mostly in single, short-lived, dramatic
fashion, we demonstrate that the nature of the transform function for $\Upsilon_e$ is not altered radically by differences in formation history. The extension of the FM over
a larger parameter space is in reality the extension of the statement that physical processes 
rather than initial
conditions or subsequent stochastic events dominate the final outcome in the formation of 
stellar systems.

\begin{acknowledgments}

DZ acknowledges financial support for this work from a
NASA LTSA award NNG05GE82G and NSF grant AST-0307482.
AIZ acknowledges financial support from NASA LTSA award NAG5-11108 and
from NSF grant AST-0206084. DZ and AIZ thank MPIA and NYU for their hospitality and
support during visits where this paper was completed. Finally, we thank 
the anonymous referee, whose comments helped us clarify various arguments.

\end{acknowledgments}

\vfill\eject

\begin{deluxetable}{lrrl}
\tabletypesize{\scriptsize}
\tablecaption{FM-Predicted Cluster Velocity Dispersions}
\tablewidth{0pt}
\tablehead{
\colhead{Name} &
\colhead{$\sigma_{PRED}$} &
\colhead{$\sigma_{OBS}$} &\colhead{Ref.} \\
&[km sec$^{-1}]$&[km sec$^{-1}]$\\}

\startdata
\\
AM1      & 0.6 & ... & ...\\
ARP2     & 0.7 & ...& ...\\
PAL3      & 1.2 & ...& ... \\
PAL4      & 1.1 & ...& ...\\
PAL5      & 0.7 & 0.9$\pm$0.2 & \cite{dehnen}\\
PAL12    & 0.6 & ... & ...\\
PAL14     & 0.5 & 0.38 $\pm$ 0.12  & \cite{jordi}\\
\enddata
\label{tb:oddballs}
\end{deluxetable} 

\begin{deluxetable}{lccccccc}
\tabletypesize{\scriptsize}
\tablecaption{$\Upsilon_*$-(log(age)) Coefficients}
\tablewidth{0pt}
\tablehead{
Order Number &
\colhead{0} &
\colhead{1} &
\colhead{2} &
\colhead{3} &
\colhead{4} &
\colhead{5} &
\colhead{6} \\
}
\startdata
\\
Coefficient &2282&$-$1736&545&$-$90.5&8.37&$-$0.409&$-$0.0082\\
\enddata
\label{tb:fitage}
\end{deluxetable} 

\begin{deluxetable}{lccc}
\tabletypesize{\scriptsize}
\tablecaption{$\Upsilon_* -(B-V)$ Coefficients}
\tablewidth{0pt}
\tablehead{
Order Number &
\colhead{0} &
\colhead{1} &
\colhead{2} \\
}
\startdata
\\
Coefficient&0.31&$-$0.55&2.2\\
\enddata
\label{tb:fitbmv}
\end{deluxetable} 
\clearpage

\clearpage
\appendix
\centerline{Appendix A}

Various efforts have been made to establish a fundamental plane for globular clusters \citep{dog, mclaughlin, barmby, pasquato}. However, 
by necessity, this plane is not the same as that for ellipticals, whose Fundamental
Plane (FP) requires $\Upsilon_e \propto \sigma^\alpha I_e^\beta$. $\Upsilon_e$ for intermediate
mass systems is already comparable to $\Upsilon_*$ and cannot physically decrease further for systems with comparable surface brightnesses but lower velocity dispersions, like globular clusters \citep{kormendy}.  

A recent study discussed a unification of all ``spheroidal" systems from globular clusters through
elliptical galaxies \citep{forbes}. While prevailing generally, they found that dwarf spheroidals
did not fit their general formalism and that the FM \citep{zzg} was not
valid down to globular clusters. We disagree with the latter statement and so we
investigate the nature of the discrepancy. 

We use their data, which they made publicly available with their paper, to remove any sample
differences. Their claim regarding the FM is based on the nature of the
empirical scaling relation for $\Upsilon_e$. In particular, they note that the globular clusters do not follow
the ``U"-shaped pattern described for $\Upsilon_e$ by \cite{zzg}. 

\begin{figure}
\epsscale{1.1}
\plottwo{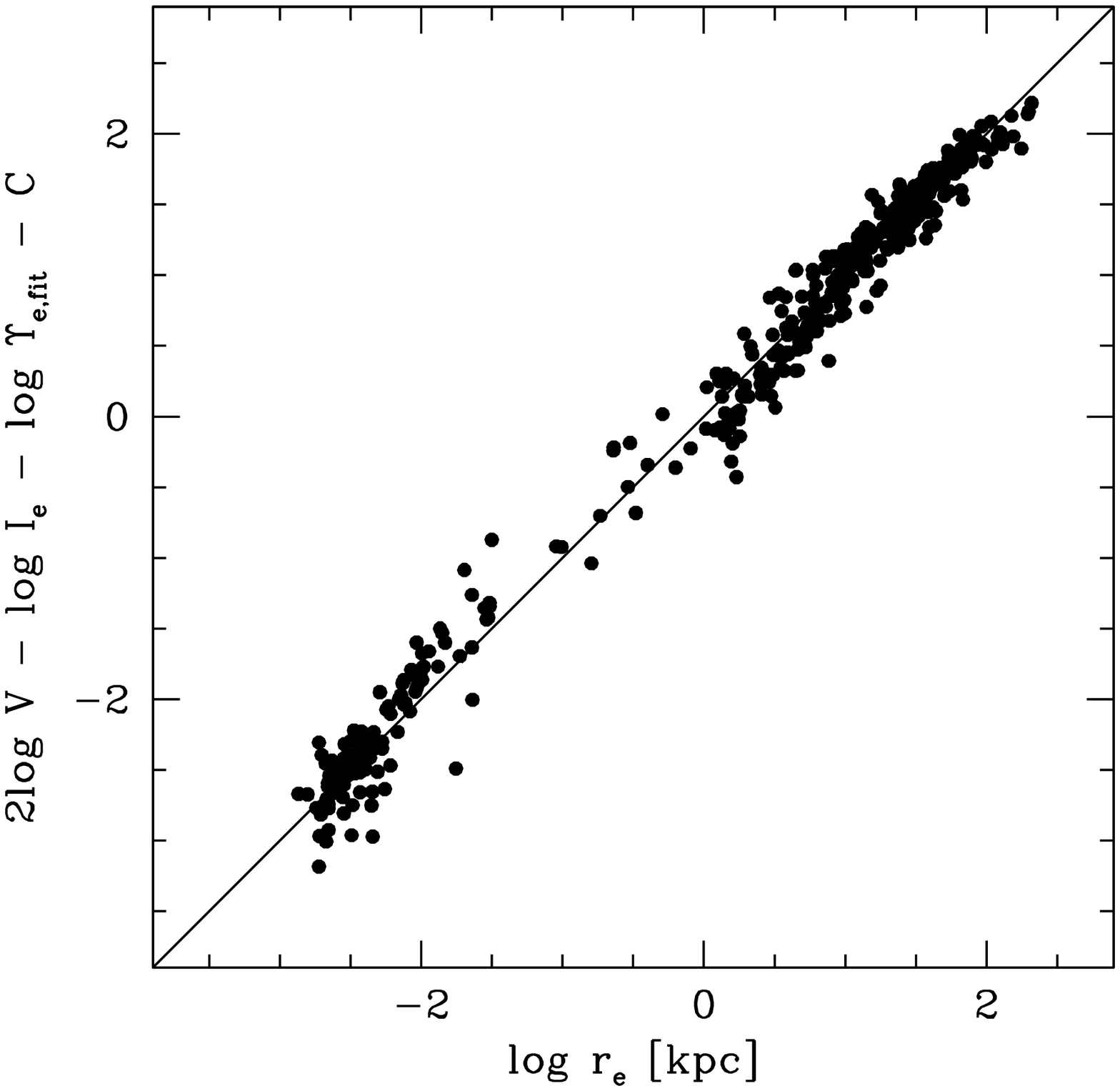}{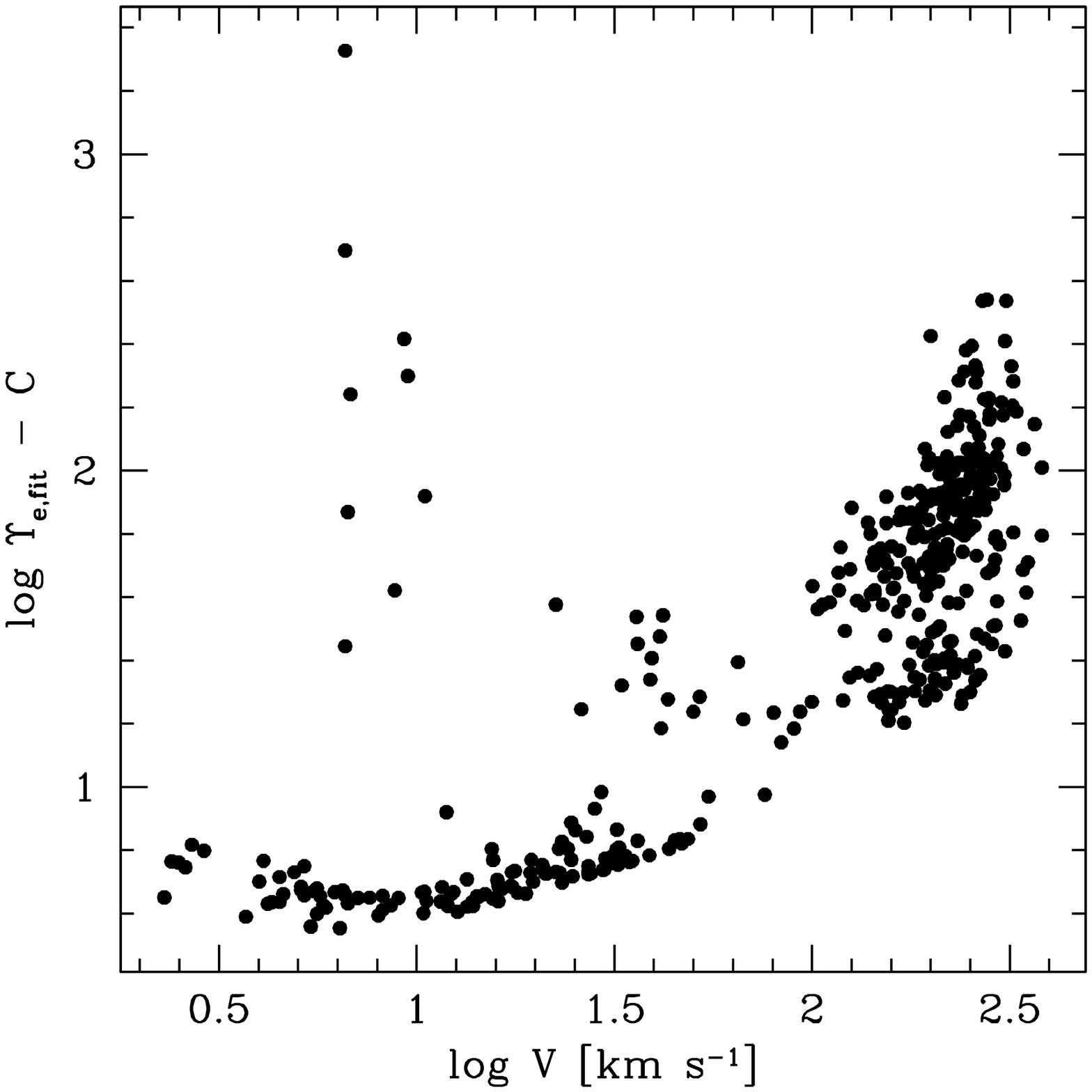}
\caption{The \cite{forbes} sample and the FM. On the left, we plot an edge-on projection of the FM
using the \cite{forbes} sample of galaxies and star clusters. We refit for the function 
$\Upsilon_{e,fit}$ because they provide $K$-band, rather than optical, data. We have not attempted to account for age effects here. That the clusters show scatter comparable to the galaxies and  are not significantly
displaced from the relationship demonstrates that this sample of star clusters does satisfy the FM even prior to accounting for possible age effects. 
On the right, we place each of their objects
on the fitting function surface to demonstrate how projections of the distribution of objects on this infinitesimally thin surface can
create large {\sl apparent} scatter. Judging whether a sample of objects satisfies the FM requires examining the ``edge-on" projection, on the left, rather than the scatter between any two parameters alone.}
\label{fig:forbes}
\end{figure}

Because their data include $K$-band magnitudes, our fitting formula for $\Upsilon_e$, which is
calibrated to $I$-band values of $\Upsilon_e$, is not optimal (color differences become large). 
Instead we refit using their data and obtain a version of $\Upsilon_{e,fit}$ appropriate for the $K$-band.
The central issue of the FM is whether the family of stellar systems can be 
described with a two-parameter relationship. The coefficients in the expansion of $\Upsilon_{e,fit}$,  or even the functional form, are
not sacrosanct, and currently have no detailed physical justification. 
The result of the new fit is shown in Figure \ref{fig:forbes}.  As \cite{zzg} stressed, the most salient feature of this Figure is not that the data straddle the 1:1 line,
because this arises by construction (although it does
imply that our functional form for $\Upsilon_{e,fit}$ is an acceptable one), but rather 
the low and uniform scatter across the full range of systems (i.e., clusters, dwarf galaxies,
and giant galaxies). 

In the right panel of Figure \ref{fig:forbes}, we explore the crux of the argument against the FM presented by \cite{forbes}.
In particular, they noted that star clusters, with low values of $\Upsilon_e$, do not follow the upward trend in $\Upsilon_e$
established by Local Group dSph galaxies. The difficulty with this argument is that it is based on a projection
of the data onto the $(\sigma, \Upsilon_e)$ axes. In the right-hand plot, we show the value of $\Upsilon_{e,fit}$ 
for each object in the \cite{forbes} sample. By construction, because we use the values on the manifold, {\sl there is zero scatter in this distribution}, and yet in this
projection there appears to be significant scatter. In fact, the clusters, which populate the lower left branch, appear
to be a distinct population from the dSphs, which lie directly above the clusters and extend upward along the $\Upsilon_e$ axis. Therefore, even points distributed on a surface with zero scatter can, in projection,
appear to comprise different populations of objects because the surface is not uniformly
populated. As such, one cannot conclude from plots such as that shown on the left of the Figure that there is no single scaling relationship among the populations.
It is therefore critical to consider the behavior of $\Upsilon_{e,fit}$ in terms of both $\sigma$ and $I_e$.
We find that the FM fits the characteristics of both high and low surface brightness objects with
low ($< 10$ km sec$^{-1}$) velocity dispersions. In particular, in our Figure \ref{fig:fit} the FM was derived excluding the star clusters and UCDs, and yet both populations fall within the uncertainties of the extrapolated relationship.

The apparent disagreement with \cite{tollerud} stems from their aims and approach. Their aim was to identify the parameter with the greatest influence in producing the full range of spheroid properties along the scaling relationship for spheroids. They correctly focused on velocity dispersion as the dominant parameter. They proceeded to derive a one-parameter scaling relationship that they could then tie to halo mass. Because they use only one parameter, they are unable to simultaneously fit the range of stellar systems with low $\sigma$ (i.e., dwarf spheroidals and star clusters/UCDs). Given their interest in tying stellar systems to dark matter halos, they naturally accepted the fit
that was consistent with the dSphs, which are clearly embedded in dark matter halos, and argued that the clusters/UCDs are a different family distinguished by their lack of dark matter halos. That is certainly one viable interpretation given the state of the data for UCDs. However, we
argue that 1) there is some evidence for UCDs having dark matter \citep{mieske}, 2)  
at the high-mass end, the UCDs overlap nicely with the low-mass end of the galaxy distribution (see Figure 1), and 3) disk galaxies also provide evidence for a two-parameter scaling relation. Our differences are therefore ones of interpretation regarding whether it is appropriate to consider all stellar systems together or to distinguish among them. Our preference for a single descriptive formalism for all stellar systems leads us in one direction, their preference for a 1:1 association of stellar systems with dark halos leads them in another. 

\begin{thebibliography}{}


\bibitem[Barmby et al.(2007)]{barmby}
{{Barmby}, P. \& {McLaughlin}, D.~E. \& {Harris}, W.~E. \& 
{Harris}, G.~L.~H. \& {Forbes}, D.~A.}, 2007, \aj, 133, 2764

\bibitem[Barnes \& Hernquist(1992)]{barnes}
Barnes, J. E., and Hernquist, L. 1991, {\sl Nature}, 360, 715 

\bibitem[Bruzual \& Charlot(2003)]{bc}
{{Bruzual}, G. \& {Charlot}, S.}, 2003, \mnras, 344, 1000

\bibitem[Burstein et al.(1997)]{burstein}
 {{Burstein}, D. \& {Bender}, R. \& {Faber}, S. \& {Nolthenius}, R.}, 1997, \aj, 114, 1365

 \bibitem[Cappellari et al.(2006)]{cappellari06}
 {Cappellari}, M.  et al. 2006, \mnras, 366, 1126.

 \bibitem[Cappellari et al.(2007)]{cappellari}
 {Cappellari}, M.  et al. 2007, \mnras, 379, 418.


\bibitem[Charbrier(2003)]{chab}
{{Chabrier}, G.}, 2003, \pasp, 115, 763

\bibitem[Conroy, Gunn, \& White(2008)]{conroy}
 {{Conroy}, C. \& {Gunn}, J.~E. \& {White}, M.}, 2008, 

\bibitem[Covington et al.(2010)]{covington}
Covington, M. et al. 2010, \apj, 710, 279

\bibitem[Dehnen et al.(2004)]{dehnen}
 {{Dehnen}, W. \& {Odenkirchen}, M. \& {Grebel}, E.~K. \& {Rix}, H.-W.},
2004, \aj, 127, 2753

\bibitem[Djorgovski(1995)]{dog}
{{Djorgovski}, S.}, 1995, \apjl, 438, L29

\bibitem[Djorgovski \& Davis(1987)]{dd87}
{{Djorgovski}, S. \& {Davis}, M.}, 1987, \apj, 313, 59

\bibitem[Dressler et al.(1987)]{d87}
{{Dressler}, A. \& {Lynden-Bell}, D. \& {Burstein}, D. \& 
{Davies}, R.~L. \& {Faber}, S.~M. \& {Terlevich}, R. \& {Wegner}, G.}, 1987, \apj, 313, 42

\bibitem[Forbes et al.(2008)]{forbes}
 {{Forbes}, D.~A. \& {Lasky}, P. \& {Graham}, A.~W. \& {Spitler}, L.}, 2008, \mnras, 389, 1924

\bibitem[Fioc \& Rocca-Volmerange(1997)]{pegase}
{{Fioc}, M. \& {Rocca-Volmerange}, B.}, 1997, \aap, 326, 950

\bibitem[Freeman \& Norris(1981)]{freeman}
{{Freeman}, K.~C. \& {Norris}, J.}, 1981, \araa, 19, 319

\bibitem[Hill \& Zaritsky(2006)]{hill}
{Hill}, A. \& {Zaritsky}, D., 2006, \aj, 131, 414

\bibitem[Hunsberger et al.(1996)]{hunsberger}
Hunsberger, S.D., Charlton, J.C., \& Zaritsky, D., 1996, \apj, 462, 50

\bibitem[Jordi et al(2009)]{jordi}
{Jordi}, K.  et al. 2009, \aj, 137, 4586

\bibitem[Kassin et al(2007)]{kassin}
Kassin, S.A., et al. 2007, 660, 35

\bibitem[Kormendy(1985)]{kormendy}
{{Kormendy}, J.}, 1985, \apj, 295, 73

\bibitem[Kroupa, Tout, \& Gilmore(1993)]{kroupa}
{{Kroupa}, P., {Tout}, C.~A., {Gilmore}, G.}, 1993, \mnras, 262, 545

\bibitem[Maraston et al.(2006)]{maraston}
{Maraston}, C. \& {Daddi}, E. \& {Renzini}, A. \& {Cimatti}, A. \& 
{Dickinson}, M. \& {Papovich}, C. \& {Pasquali}, A. \& {Pirzkal}, N.,
2006, \apj, 652, 85

\bibitem[McLaughlin(2000)]{mclaughlin}
{{McLaughlin}, D.~E.}, 2000, \apj, 539, 618

\bibitem[McLaughlin \& van der Marel(2005)]{clusters}
{{McLaughlin}, D.~E. \& {van der Marel}, R.~P.}, 2005, \apjs, 161, 304

\bibitem[Mieske et al.(2008)]{mieske}
Mieske, S. et al. 2008, A\&A, 487, 921

\bibitem[Paquato \& Bertin(2008)]{pasquato}
{{Pasquato}, M. \& {Bertin}, G.}, 2008, \aap, 489, 1079

\bibitem[Reijns et al.(2006)]{omegacen}
{Reijns}, R.~A. \& {Seitzer}, P. \& {Arnold}, R. \& {Freeman}, K.~C. \& {Ingerson}, T. \& {van den Bosch}, R.~C.~E. \& {van de Ven}, G. \& {de Zeeuw}, P.~T., 2006, \apj, 445, 503

\bibitem[Tollerud et al.(2010)]{tollerud}
 {{Tollerud}, E. \& {Bullock}, J.~S. \& {Graves}, G.~J., {Wolf}, J., {Barton}, E.J.},
2010, arXiv:astro-ph/0501272

\bibitem[Tonini et al.(2008)]{tonini}
{{Tonini}, C., {Maraston}, C.,{Devriendt}, J.,{Thomas}, D., \&
{Silk}, J.}, 2008, ArXiv/0812.1225

\bibitem[Walsh et al.(2008)]{walsh}
Walsh, S.M. et al. 2008, \apj, 688,245

\bibitem[Weiner et al.(2006)]{weiner06}
Weiner, B. J., et al. 2006, \apj, 653, 1049

\bibitem[Zaritsky  et al.(2006a)]{zgz}
 {{Zaritsky}, D. \& {Gonzalez}, A.~H. \& {Zabludoff}, A.~I.}, 2006, \apj, 638, 725

\bibitem[Zaritsky et al.(2006b)]{zgzb}
{{Zaritsky}, D. \& {Gonzalez}, A.~H. \& {Zabludoff}, A.~I.}, 2006, \apjl, 642, L37

\bibitem[Zaritsky et al.(2008)]{zzg}
 {{Zaritsky}, D. \& {Zabludoff}, A.~I. \& {Gonzalez}, A.~H.}, 2008, \apj, 682, 68

\end{thebibliography}
\end{document}